\documentclass[epj]{svjour}

\usepackage{latexsym}
\usepackage{amsmath}
\usepackage{amsfonts}
\usepackage{graphicx}
\newcommand{\kln}{ \ln_{_{\{\kappa\}}} }
\newcommand{\kexp}{ \exp_{_{\{\kappa\}}} }
\newcommand{\ku}{ u_{_{\{\kappa\}}} }

\begin{document}
\title{Asymptotic solutions of a nonlinear diffusive equation in the framework of $\kappa$-generalized statistical mechanics}

\titlerunning{Asymptotic solutions of the $\kappa$-generalized diffusive equation}
\author{T. Wada\inst{1} \and A.M. Scarfone\inst{2}% etc
}                     % Do not remove
\institute{Department of Electrical and Electronic
Engineering,
Ibaraki University, Hitachi,~Ibaraki, 316-8511, Japan.\\
\email{wada@mx.ibaraki.ac.jp} \and Istituto Nazionale di
Fisica della Materia (CNR-INFM) and Dipartimento di Fisica,
Politecnico di Torino, I-10129, Italy.\\
\email{antonio.scarfone@polito.it}}
\authorrunning{T. Wada and A.M. Scarfone}

\date{Received: date / Revised version: date}

%% Please type your abstract here.
\abstract
{The asymptotic behavior of a nonlinear diffusive equation
obtained in the framework of the $\kappa$-generalized statistical mechanics is studied.
The analysis based on the classical Lie symmetry shows that the $\kappa$-Gaussian function is not a scale invariant solution of the generalized diffusive equation.
Notwithstanding, several numerical simulations, with different
initial conditions, show that the solutions asymptotically approach to the
$\kappa$-Gaussian function.
Simple argument based on a time-dependent transformation performed on the related
$\kappa$-generalized Fokker-Planck equation, supports this conclusion.
\PACS{
      {05.20.Dd}{Kinetic theory} \and
      {05.20.-y}{Classical statistical mechanics}   \and
      {05.90.+m}{Other topics in statistical physics}
     } % end of PACS codes
} %end of abstract
\maketitle
%

%% Keywords should be separated by \*\ sign
%\keywords{nonlinear diffusive equations \*\ $\kappa$-entropy \*\
% asymptotic behavior \*\ Lie symmetric analysis}

%% ###################################################################

\section{Introduction}
In the framework of non-equilibrium thermostatistics, irreversible processes can be often described
by means of Fokker-Planck equations (FPE's) \cite{Frank} whose time evolutions
are characterized by monotonically non-increasing Lyapunov functionals.
In a previous work \cite{WS07}, we
derived a non-linear FPE in the picture of a generalized statistical mechanics based on
the $\kappa$-entropy \cite{k-entropy1,k-entropy2}, and discussed its relation with
the associate Lyapunov functional or Bregman type divergence \cite{Bregman}.
Based on the monotonic behavior of the Lyapunov functional, one can show that
any initial localized state, which evolves according to
the $\kappa$-generalized FPE, converges to a stationary solution which
minimizes the Lyapunov functional.
In addition, in the linear drift case, the corresponding
stationary solution is nothing but the $\kappa$-Gaussian, which is
a generalization of Gaussian by replacing the standard exponential
with its $\kappa$-generalized version \cite{WS06}.\\
As well known, the diffusive equations play a fundamental r\^{o}le in
the description of several physical phenomena. In particular, the linear diffusive equation arising in the Boltzmann-Gibbs (BG) statistical mechanics has been widely studied jointly with its Gaussian self-similar solution \cite{Risken}.
Similarly, the $q$-generalized diffusive equation arises, in a natural way,
in the framework of the Tsallis statistical physics.
This is equivalent to the porous medium equation (PME),
which is a transport equation of gases or fluids in a porous medium.
The main properties of PME are now well known \cite{PME}
and its solutions have been widely investigated in literature \cite{Carrillo,Witelski}, in particular
its self-similar solution is given by the so-called Barenblatt
solution \cite{Barenblatt}.\\
In this work we study a different nonlinear diffusive equation obtained in the picture of the
$\kappa$-generalized statistical mechanics.
As known, a useful method to obtain the solutions of
a partial differential equation is
based on the study of its Lie symmetries and the related group invariant solutions \cite{Olver,Hydon}. We accomplished such analysis for the $\kappa$-diffusive equation considering only the classical Lie symmetries whose generators are functions of the independent and the dependent variables. We will leave the study of the generalized Lie symmetries, whose generators are functions also of the derivatives of the depending variables, to a future work.\\
In particular, we show that the only
group-invariant solutions for the $\kappa$-diffusive equation are the kink-like solutions (a non-normalizable family of self-similar solutions) and the traveling wave solutions (physically irrelevant because they are divergent).
Next, we explore numerically the evolutions of a localized initial
state, whose spreading is governed  by the $\kappa$-diffusive equation.
It is shown that, independently from the initial states, the numerical solutions
approach to a shape which is asymptotically well fitted by the $\kappa$-Gaussian distribution, although this one is not a group invariant solution of the equation under investigation.

The paper is organized as follows.
In the next section 2 we briefly review the $\kappa$-generalized thermostatistics
and its associated non-linear FPE.
In section 3, after presenting an alternative derivation of the $\kappa$-diffusive equation, we classify its classical Lie symmetries and their related group invariant
solutions. Section 4 deals with the numerical analysis and the final section 5 is devoted to summary.

%%%%%%%%%%%%%%%%%%%%%%%%%%%%%%%%%%%%%%%%%%%%%%%%%%%%%%%%%%%%%%%%%%%%%%%%%%%%%%%%%%%%%%%%

\section{$\kappa$-generalized thermostatistics}
We briefly review the generalized thermostatistics based on the
$\kappa$-entropy \cite{k-entropy1,k-entropy2} given by
\begin{equation}
  S_{\kappa}[p] \equiv -\int\limits_{-\infty}\limits^{\infty}  p(v) \kln p(v)\,dv \ ,\label{kent}
\end{equation}
where $\kln(x)$ is the $\kappa$-logarithm defined us
\begin{equation}
  \kln(x) = \frac{x^{\kappa} - x^{-\kappa}}{2 \kappa} \ .
  \label{k-log}
\end{equation}
The $\kappa$-entropy $S_{\kappa}[p]=S_{-\kappa}[p]$ is an extension of the BG entropy
by means of a real parameter $|\kappa|\in[0,\,1]$.\\
The inverse function of $\kln(x)$, namely $\kexp(x)$, given by
\begin{equation}
\kexp(x) = \left( \kappa x + \sqrt{1 + \kappa^2 x^2}
       \right)^{\frac{1}{\kappa}} \ ,
\end{equation}
is called $\kappa$-exponential.\\
The $\kappa$-logarithm and $\kappa$-exponential functions are
the building blocks of the statistical mechanics based on the $\kappa$-entropy. In particular, the $\kappa$-exponential is a monotonic increasing and convex function, fulfilling the relationship $\kexp(-x)=1/\kexp(x)$. Moreover,
$\kexp(-\infty)=0$, $\kexp(0)=1$ and $\kexp(\infty)=\infty$, as the ordinary exponential function does.
For $|x|\ll1$, the $\kappa$-exponential
is well approximated by the standard exponential, whereas for $|x|\to\infty$,
it asymptotically approaches to a power-law: $\kexp(x) \sim |2\,\kappa\,x|^{\pm1/\kappa}$.
In the $\kappa\to0$ limit, both $\kln(x)$ and $\kexp(x)$ reduce to
the standard logarithm and exponential functions, respectively.
Accordingly, the $\kappa$-entropy reduces to the BG entropy.

Maximizing $S_{\kappa}[p]$ under the constraints of
the linear kinetic energy average and the normalization
\begin{equation}
   \frac{\delta}{\delta p(v)} \Big(
     S_{\kappa}[p]-\beta \int\limits_{-\infty}\limits^{\infty} \frac{1}{2}\,v^2\,p(v)\,dv-\gamma \int\limits_{-\infty}\limits^{\infty} p(v)\,dv \Big) = 0 \ ,
\end{equation}
leads to the $\kappa$-Gaussian function
\begin{equation}
 p^{\rm ME}(v) = \alpha \,\kexp
  \left(- \frac{1}{\lambda} \Big(\gamma + \beta \, \frac{v^2}{2} \Big)\right) \ .\label{kg}
\end{equation}
Here $\gamma$ is the constant fixing the normalization of the distribution and depends
on the Lagrange multiplier $\beta$ which controls the dispersion of the distribution.
Finally, the parameters $\alpha$ and $\lambda$ are $\kappa$-dependent constants given by
\begin{equation}
 \alpha = \left(\frac{1-\kappa}{1+\kappa}\right)^{\frac{1}{2\kappa}} \ , \quad
 \lambda = \sqrt{1-\kappa^2} \ ,
\end{equation}
respectively, and are related each to the other through the relation
\begin{equation}
  \alpha  = \kexp\left( -1/\lambda \right) \ .
\end{equation}

The $\kappa$-entropy and the corresponding statistical mechanics preserve
several properties of the BG theory \cite{k-entropy1}.
In particular, the Legendre structure of the thermo-statistics theory has been investigated in \cite{SW06} through the introduction of several thermodynamic potentials.\\
For instance, it is found that the generalized partition function,
as a function of $\beta$, given by
\begin{equation}
\kln\left(Z_\kappa\right)={\cal I}_\kappa[p]+\gamma \ ,\label{Z}
\end{equation}
satisfies the Legendre relation
\begin{equation}
{d\over d\beta}\kln\left(Z_\kappa\right)=-U,
\end{equation}
and that the $\kappa$-entropy and the generalized partition
function are related through the relationship
\begin{equation}
S_\kappa=\kln\left(Z_\kappa\right)+\beta\,U \ ,
\end{equation}
where
\begin{equation}
U=\int\limits_{-\infty}\limits^{\infty} \frac{1}{2}\,v^2\,p(v)\,dv \ .
\end{equation}
The function ${\cal I}_\kappa[p]$ in equation (\ref{Z}) is defined by
\begin{equation}
 {\cal I}_{\kappa}[p]\equiv \int\limits_{-\infty}\limits^{\infty}p(v)\,\ku\big(p(v)\big)\,dv \ ,
\end{equation}
where the quantity $\ku(x)=u_{_{\{-\kappa\}}}(x)$, defined as
\begin{equation}
\ku(x)=\frac{x^\kappa+x^{-\kappa}}{2} \ ,
\end{equation}
fulfills the following proprieties $\ku(x)=\ku(1/x)$ and $\ku(\alpha)=1/\lambda$. In the $\kappa\to0$ limit, it reduces to the unity: $u_{_{\{\kappa\}}}(x)=1$.\\ Both functions $\ku(x)$ and ${\cal I}_{\kappa}[p]$ play an important r\^ole in the development of a theory based on the $\kappa$-entropy.\\
Starting from the definition of the partition function $Z_\kappa$, we can introduce a $\kappa$-generalization of the free-energy, according to
\begin{equation}
  F_{\kappa}[p] \equiv -{1\over\beta}\kln\left(Z_\kappa\right) \ ,
\end{equation}
In this way, it was shown that the $\kappa$-free-energy satisfies the Legendre transformation
structures \cite{SW06} summarized by the following relationships
\begin{equation}
F_{\kappa}[p] = U[p] - \frac{1}{\beta} \, S_{\kappa}[p] \ , \quad
   \frac{d}{d \beta} \, \Big( \beta\,F_{\kappa} \Big) = U \ .
\end{equation}

\subsection{$\kappa$-generalized Fokker-Planck equation}
In a previous work \cite{WS07}, we have studied the nonlinear FPE
associated with the $\kappa$-entropy. It can be introduced as a continuity
equation for the density field $p(v,t)$
\begin{equation}
\frac{\partial}{\partial t} p(v, t) =\frac{\partial}{\partial v}j(v,t) \ ,
\end{equation}
with the current density
\begin{equation}
j(v,t)=-p(v,t)\frac{\partial}{\partial v}\left(\frac{\delta{\cal L}_\kappa[p]}{\delta p(v,t)}\right) \ ,
\end{equation}
and
\begin{equation}
  {\mathcal L}_{\kappa}[p] \equiv U[p] - D \, S_{\kappa}[p] \ .
\end{equation}
As shown in \cite{WS07} the quantity ${\cal L}_\kappa[p]$ is a $\kappa$-generalized Lyapunov functional,
which is monotonically non-increasing in time.\\
The $\kappa$-generalized FPE reads
\begin{equation}
  \frac{\partial}{\partial t} p(v, t) =
  \frac{\partial}{\partial v} \Big( v \, p(v, t) \Big)
  + D \frac{\partial^2}{\partial v^2}
   \Big( p(v,t)\,\ku\big(p(v,t)\big) \Big) \ ,
\label{kFPE}
\end{equation}
where $D$ is a constant diffusion coefficient.\\ We remark that the current density $j(v, t)=j^{\rm drift}(v, t)+j^{\rm diff}(v, t)$ is the sum of a linear drift current $j^{\rm drift}(v, t)=v\,p$, which describe the standard Uhlenbeck-Ornstein process and a nonlinear diffusive current, given by $j^{\rm diff}(v, t)=D\,\partial\big[p\,\ku(p)\big]/\partial v$, which reduces to the standard Fick's current $j^{\rm Fick}(v, t)=D\,\partial p/\partial v$ in the $\kappa\to0$ limit. Consequently, in the same limit, equation (\ref{kFPE}) reduces to the standard linear FPE
\begin{equation}
\frac{\partial}{\partial t} p(v, t) =
  \frac{\partial}{\partial v} \Big( v \, p(v, t) \Big)
  + D \frac{\partial^2}{\partial v^2} p(v,t) \ .
\end{equation}
Since $d{\mathcal L}_{\kappa}/dt \le 0$,
the functional ${\mathcal L}_{\kappa}[p]$ is minimized for the stationary solutions of equation (\ref{kFPE}) according to
\begin{align}
 \min {\mathcal L}_{\kappa}[p]
 &= \lim_{t \to \infty} {\mathcal L}_{\kappa}[p] \nonumber \\
 &\equiv U[p^{\rm st}] - D \, S_{\kappa}[p^{\rm st}] = F_{\kappa} [p^{\rm st}] \ .
\end{align}
It is easy to verify that the stationary solution of the $\kappa$-generalized FPE
is equal to the optimal maximal entropy distribution $p^{\rm ME}(v)$, i.e.
\begin{align}
p^{\rm st}(v) &\equiv \lim_{t \to \infty} p(v, t) \\
   &=  \alpha \,\kexp
  \left(- \frac{1}{\lambda} \Big(\gamma + \frac{1}{D} \,
  \frac{v^2}{2} \Big)\right)  = p^{\rm ME}(v),
  \label{stationary}
\end{align}
with the position $D=1/\beta$.\\
Without the external potential $U$ the drift therm $j^{\rm drift}$ disappear and equation (\ref{kFPE}) reduces to a purely $\kappa$-diffusive equation whose explicit form is given by
\begin{equation}
  \frac{\partial}{\partial t} p(v, t) =
  D \frac{\partial^2}{\partial v^2}
   \left( \frac{ p(v,t)^{1+\kappa} + p(v,t)^{1-\kappa}}{2} \right) \ .
  \label{kDE}
\end{equation}
Equation (\ref{kDE}) is the main subject of our investigation.

%%%%%%%%%%%%%%%%%%%%%%%%%%%%%%%%%%%%%%%%%%%%%%%%%%%%%%%%%%%%%%%%%%%%%%%%%%%%%%%%%%%%%%%%

\section{$\kappa$-generalized diffusive equation}

\subsection{A simple physical derivation}
First, let us describe a physical derivation of the $\kappa$-diffusive equation (\ref{kDE}).
In this respect, we recall that the linear diffusive equation 
\begin{equation}
  \frac{\partial}{\partial t} p(v, t) =
  D \frac{\partial^2}{\partial v^2}
p(v,t) \ ,
\end{equation}
can be
obtained from the following physical relations:
\begin{align}
  \left\{
\begin{array}{ll}
  \frac{\partial}{\partial t} \rho(v,t) + \frac{\partial}{\partial v}\big(v\,\rho(v,t)\big)=0
  & \textrm{(continuity equation)} \\
  v \propto - \frac{\partial P}{\partial v} & \textrm{(Darcy's law)} \\
  P \propto \ln(\rho(v,t)\big) & \textrm{(equation of state)} \\
\end{array}
\right.
\end{align}
where $v$ denotes the average velocity
of the fluid and $P$ denotes the pressure of the fluid, which is related to the density of the fluid $\rho(v,t)$ through a suitable equation of state $P\propto\Phi(p)$, with $\Phi(x)=\ln(x)$.\\
The continuity equation assures the conservation of the zero-th moment
\begin{equation}
 \int \rho(v, t) dv = M \ ,
\end{equation}
which corresponds to the total mass $M$ of the fluid.

The PME \cite{PME} can be derived from the same physical relations by posing, in the equation of state, $\Phi(x)=x^\nu$:
\begin{align}
  \left\{
\begin{array}{ll}
  \frac{\partial}{\partial t} \rho(v,t) + \frac{\partial}{\partial v}\big(v\,\rho(v,t)\big)=0
  & \textrm{(continuity equation)} \\
  v \propto - \frac{\partial P}{\partial v} & \textrm{(Darcy's law)} \\
  P \propto \rho(v,t)^{\nu} & \textrm{(polytropic fluid)} \\
\end{array}
\right.
\end{align}
Here, $\nu$ is a positive real parameter characterizing the degree
of porosity of the medium and
characterizing also the power-law dependence of polytropic fluid.
In other words, $\nu$ is a polytropic index.\\
The corresponding PME is given by
\begin{equation}
 \frac{\partial}{\partial t} \rho(v, t)
  = D \, \frac{\partial^2}{\partial v^2}\, \rho(v, t)^{q} \ ,
\label{PME}
\end{equation}
with $q = \nu + 1$.\\
Interestingly, the same settings can be also applied to the $\kappa$-diffusive equation (\ref{kDE}), merely replacing, in the equation of state, the logarithm with its $\kappa$-generalization: $\Phi(x)=\kln(x)$.\\ In this way, the $\kappa$-diffusive equation (\ref{kDE}) can be derived according to
\begin{align}
  \left\{
\begin{array}{ll}
  \frac{\partial}{\partial t} \rho(v,t) + \frac{\partial}{\partial v}\big(v\,\rho(v,t)\big)=0
  & \textrm{(continuity equation)} \\
  v \propto - \frac{\partial P}{\partial v} & \textrm{(Darcy's law)} \\
  P \propto \kln\big(\rho(v,t)\big) &
\textrm{(a special type of barotropic fluid)} \\
\end{array}
\right.
\end{align}
which describe a special type of \textit{barotropic} fluid \cite{barotropic}.
``A special type'' means as follows.
From the definition of $\kappa$-logarithm \eqref{k-log}, the last relation
can be written as
\begin{align}
   P \propto (\rho^{\kappa} - \rho^{-\kappa}),
\end{align}
i.e., a pressure $P$ of this type of barotropic fluid depends
on the difference between one term with a polytropic index $\kappa$
and the other term with $-\kappa$.

\subsection{Classical Lie symmetry analysis}
In the following we classify the Lie symmetries of the $\kappa$-diffusive equation.\\
For the sake of comparison, we first consider the PME \eqref{PME}. In this case, the symmetry generators are given by (please refer to \cite{Bluman} for more details)
\begin{align}
   &v^{(1)} = {\partial\over\partial t} \ ,\label{shift1}\\
   &v^{(2)} = {\partial\over\partial v} \ ,\label{shift2} \\
   &v^{(3)} = v\,{\partial\over\partial v}+2\,t\,{\partial\over\partial t} \ ,\label{scaling1}\\
   &v^{(4)} = p\,{\partial\over\partial p}+(1-\nu)\,t\,{\partial\over\partial t} \ .\label{scaling2}
\end{align}
This means that if $p(v, t)$ is a solution of PME, then
\begin{align}
   &p^{(1)}(v,t) = p(v, t-\epsilon) \ ,\label{vshift}\\
   &p^{(2)}(v,t) = p(v-\epsilon, t) \ ,\label{tshift} \\
   &p^{(3)}(v,t) = p(e^{-\epsilon} v, e^{-2 \epsilon} t) \ ,\label{scaling3}\\
   &p^{(4)}(v,t) = e^{-\epsilon} p(v, e^{(\nu-1) \epsilon} t) \ ,\label{scaling4}
\end{align}
are also solutions.\\
Generators (\ref{shift1}) and (\ref{shift2}) describe translations invariance in $x$- and $t$-coordinates, respectively and their group invariant solutions (GIS) are merely the constant functions.\\
Generators (\ref{scaling1}) and (\ref{scaling2}) describe scaling invariance of independent and dependent variables, respectively. In particular, the GIS for the symmetry $v^{(3)}$ has a kink-like shape and in the $q\to1$ limit, it reduces to the Erf$(x)$ function.\\
Further, GIS can be obtained by means of linear combination of 
the vectors $v^{(i)} (i=1,2,3,4)$. For instance, from the generator $v=v^{(1)}+a\,v^{(2)}$ we obtain the traveling wave solutions which are physically irrelevant because they diverge. Differently, the most general self-similar generator is provided by vector $v=v^{(3)}+a\,v^{(4)}$. Although the corresponding GIS can not be given in terms of elementary functions, explicit solutions can be obtained for special values of the constant $a$. In particular, for $a=-1$ we derive
the Barenblatt scale invariant solution \cite{Barenblatt}
\begin{equation}
 p^{\rm B}(v, t) = (D \,t)^{-\frac{1}{\mu+1}}
  \Big( C - \frac{\mu-1}{2 \mu(\mu+1)} v^2 (D \,t)^{-\frac{2}{\mu+1}}
         \Big)^{\frac{1}{\mu-1}},
\end{equation}
where $C$ is the normalization constant. Note that this solution is nothing but a $q$-Gaussian with $\mu=2-q$ \cite{gq}.
\begin{figure}
\begin{center}
\includegraphics[width=0.42\textwidth]{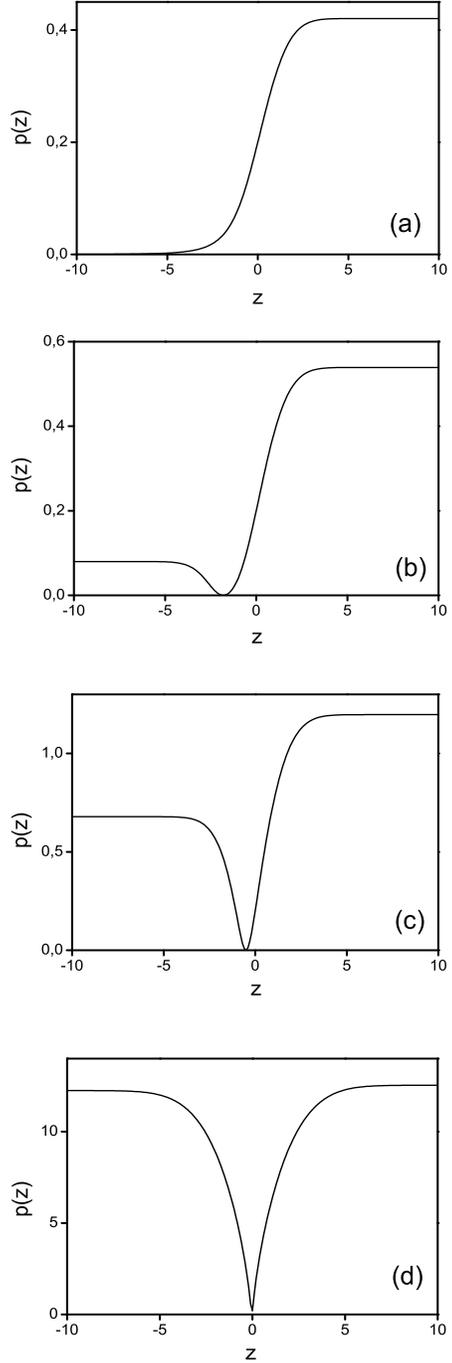}
\caption{Kink-like self-similar solutions of the $\kappa$-diffusion equation $p(x,t)\equiv p(z)$ with $z=v^2/t$,
for $\kappa=0.5$ and several boundary conditions: (a) $p(0)=0.2$ and $p'(0)=0.13$; (b) $p(0)=0.2$ and $p'(0)=0.2$; (c) $p(0)=0.2$ and $p'(0)=0.6$; (d) $p(0)=0.2$ and $p'(0)=10$.
 \label{fig1}}
\end{center}
\end{figure}
Next, we study the symmetries for the more complicated $\kappa$-diffusive equation. It is worth noting that the PME is characterized by the single-diffusive term $\rho^q$, which leads to the GIS \eqref{scaling4},
while the $\kappa$-diffusive equation contains two different powers $p^{1+\kappa}$
and $p^{1-\kappa}$.
This double nonlinearity in the diffusive term destroys the scaling invariance for the dependent variable $p(v,t)$.\\ Thus, the only Lie symmetry generators are
\begin{align}
   &v^{(1)} = {\partial\over\partial t} \ ,\label{shift1a}\\
   &v^{(2)} = {\partial\over\partial v} \ ,\label{shift2a} \\
   &v^{(3)} = v\,{\partial\over\partial v}+2\,t\,{\partial\over\partial t} \ ,\label{scaling1a}
\end{align}
so that, if $p(v,t)$ is a solution of the $\kappa$-generalized
diffusive equation then
\begin{align}
   p^{(1)} &= p(v, t-\epsilon) \ ,\\
   p^{(2)} &= p(v-\epsilon, t) \ ,\\
   p^{(3)} &= p(e^{-\epsilon} v, e^{-2 \epsilon} t) \ ,
\end{align}
are also solutions.\\ Again, the GIS corresponding to the traveling wave generator $v=v^{(1)}+a\,v^{(2)}$ are physically irrelevant because merely constants or divergent.
Differently, the only self-similar generator is $v^{(3)}$ and the corresponding GIS has a kink-like shape as shown in figure \ref{fig1}.
These solutions, although not normalizable, can describe the evolution of an order parameter belonging to the system. However, as compared to the case of PME, there is no scaling generator
like equation \eqref{scaling2} and consequently, the $\kappa$-Gaussian is not an exact solution for
the $\kappa$-diffusive equation.
%%%%%%%%%%%%%%%%%%%%%%%%%%%%%%%%%%%%%%%%%%%%%%%%%%%%%%%%%%%%%%%%%%%%%%%%%%%%%%%%%%%%%%%%

\section{Asymptotical analysis}
\subsection{Numerical simulations}
\begin{figure}
\begin{center}
\includegraphics[width=0.4\textwidth]{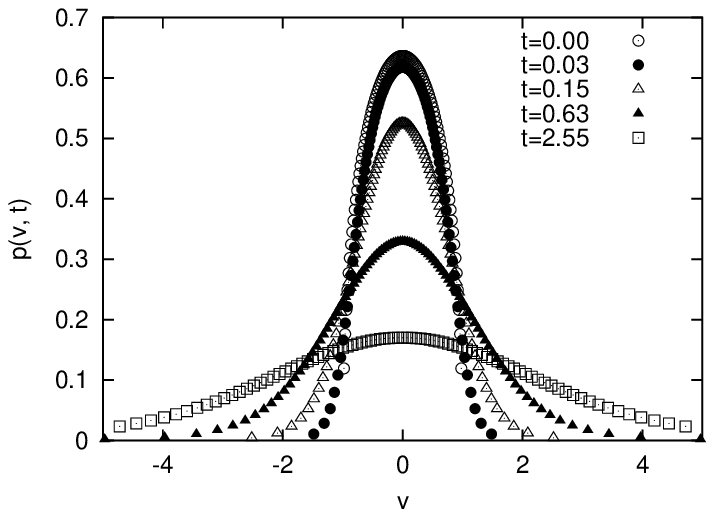}
\includegraphics[width=0.4\textwidth]{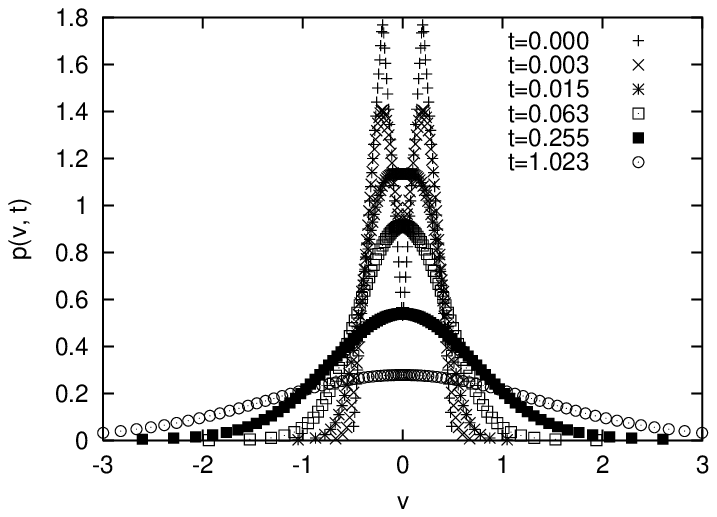}
\caption{Top: a typical time evolution for the $\kappa$-diffusion equation
with $\kappa=0.3$. Bottom: The same for an initial probability
distribution function with double peaks.
 \label{typ-evo}}
\end{center}
\end{figure}

\begin{figure}
\begin{center}
\includegraphics[width=0.33\textwidth]{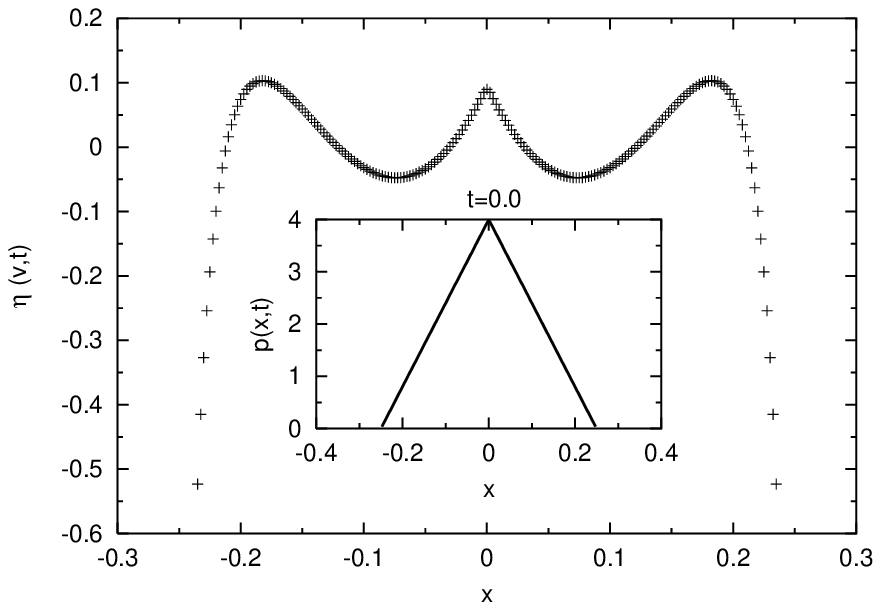}
\includegraphics[width=0.33\textwidth]{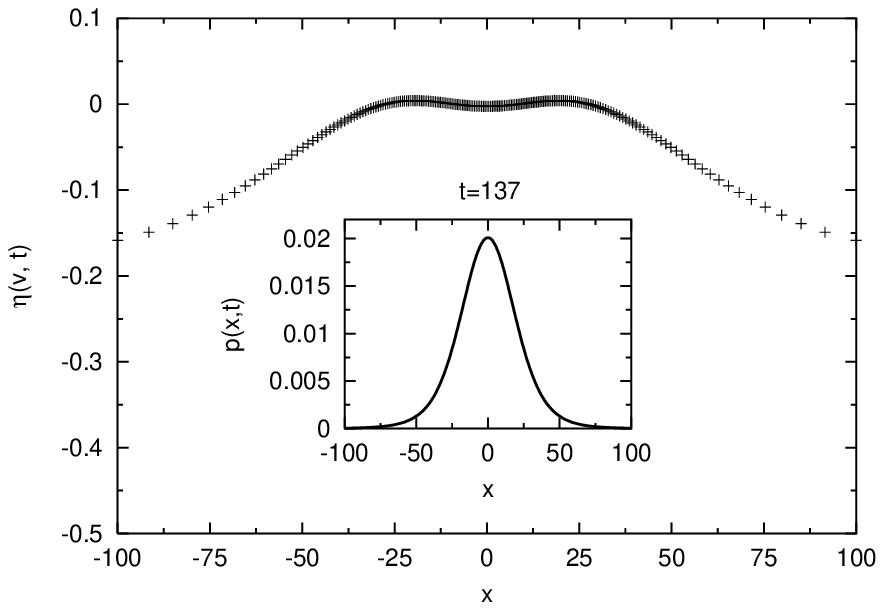}
\includegraphics[width=0.33\textwidth]{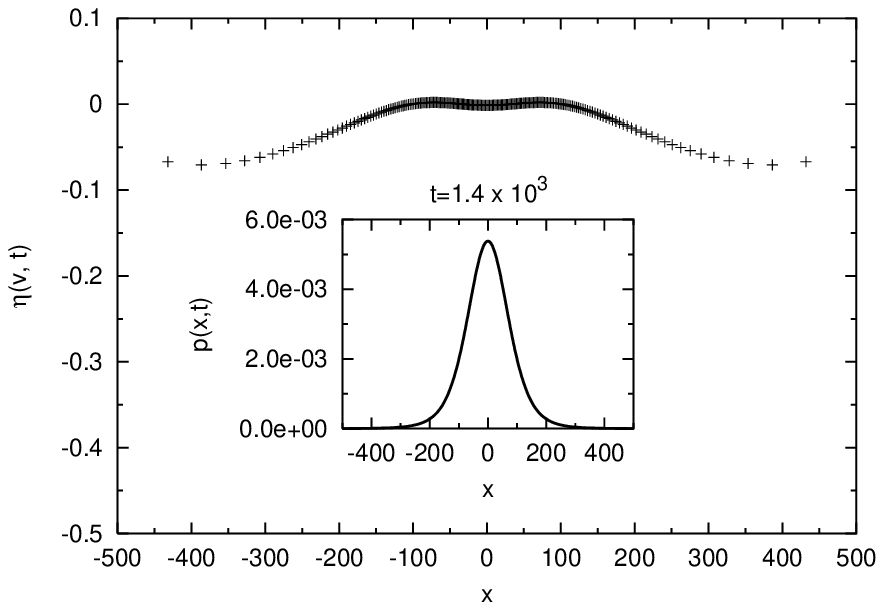}
\includegraphics[width=0.33\textwidth]{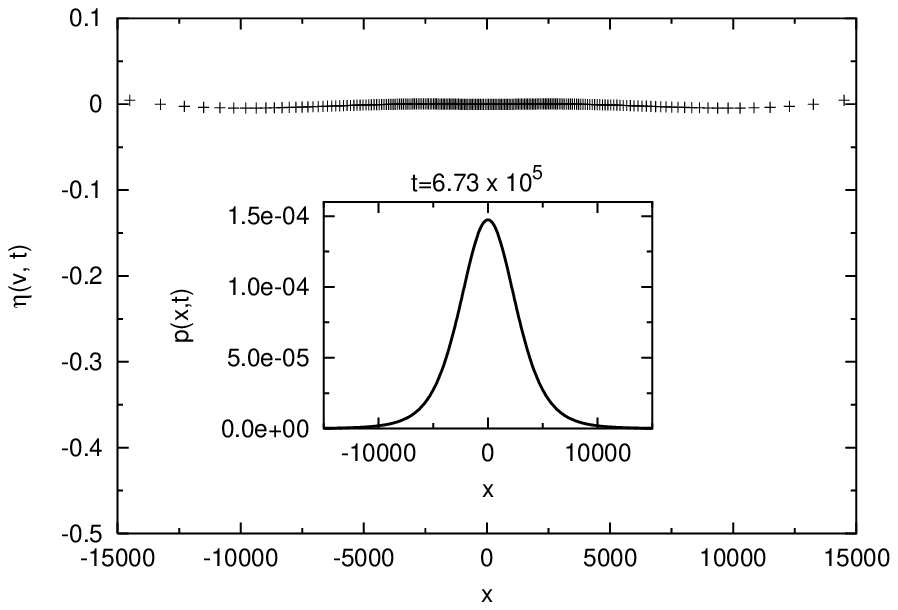}
\caption{Long time behavior of the numerical solutions with
an initial triangular probability distribution.
The parameter $\kappa=0.3$
and the number of calculated points for $\bar{r}$ are $201$.
The quantity $\eta(v,t)$ of equation \eqref{ratio} is plotted in each figure.
The best fitted parameters in \eqref{k-Gau} are: $a=1.324, b=40.54$ at $t=0.0$;
$a=-4.862, b=2.777 \times 10^{-3}$ at $t=137$;
$a=-7.640, b=2.871 \times 10^{-4}$ at $t=1.4 \times 10^3$
and $a=-23.39, b=6.205 \times 10^{-7}$ at $t=6.73 \times 10^5$, respectively.
Inset figures show $p(v,t)$ at each time $t$.
\label{logratio}}
\end{center}
\end{figure}

To further study the long time behavior of the $\kappa$-diffusion,
we performed numerical simulations against various initial probability
distribution functions with different shapes.\\
We used a numerical method developed in \cite{GT06}
in order to study the approach to the equilibrium of a given Cauchy problem for the one-dimensional generalized diffusive equation
\begin{equation}
 \frac{\partial}{\partial t} \, p(v, t) = \frac{\partial^2}{\partial v^2}
  \Phi\Big( p(v,t) \Big), \quad p(v, t=0) \equiv p_0(v) \ge 0 \ ,
  \label{g-PME}
\end{equation}
where $p_0(v) \in L^1(R)$ is an initial localized state and
$\Phi(p) \in C^2(\mathbb{R}^+)$.
For computational simplicity, the diffusion coefficient $D$ is re-scaled in the independent variables according to the symmetry (\ref{scaling1a}), which still holds whatever the function $\Phi(p)$ is.\\
In general, a solution $p(v,t)$ of any diffusive equation spreads
out as time increases. Then, simple conventional
methods of discretizations in the $v$-coordinate, which are often employed in the numerical simulations, leads to much computational costs, especially for a long time numerical simulation.\\
In order to overcome the difficulty due especially to the expanding support of $p(v,t)$,
Gosse and Toscani \cite{GT06} developed an explicit numerical scheme applicable to one-dimensional
differential equations of the type given in \eqref{g-PME}.\\
We here explain the basic points of this numerical scheme.
First, we introduce the cumulative distribution function
\begin{equation}
    r(v, t) \equiv \int_{-\infty}^{v} p(\xi, t) d\xi \in [0, 1]\ ,
\end{equation}
associated with the probability distribution $p(v,t)$ and its inverse relation
\begin{equation}
  \frac{\partial}{\partial v} \, r(v,t) = p(v,t) \ .
\end{equation}
By using equation \eqref{g-PME} and integrating by part we obtain
\begin{align}
   \frac{\partial}{\partial t} \, r(v,t) &=
\int_{-\infty}^v \frac{\partial}{\partial t}\, p(\xi,t) d\xi
 =\int_{-\infty}^v \frac{\partial^2 }{\partial \xi^2}\,
 \Phi\left( p(\xi,t) \right) d\xi \nonumber \\
  & = \frac{\partial}{\partial v}
    \Phi\left( \frac{\partial}{\partial v} r(v,t) \right) \ .
\end{align}
Next, since $r(v,t)$ is non-decreasing
in the $v$ variable, one can define its pseudo-inverse function for
a given value $\bar{r} \in [0,1]$ as
\begin{equation}
    v(\bar{r},t) \equiv \inf \left\{ v \in \mathbb{R},
    \textrm{ such that } r(v, t) = \bar{r} \right\} \ .
\end{equation}
Then, we have
\begin{equation}
    r\left( v(\bar{r}, t),t \right) = \bar{r} \ .\label{r}
\end{equation}
Note that $\bar{r}$ does not depend on time $t$. In other words,
at each time $t$ there exists only one point of $v(\bar{r},t)$
whose cumulative distribution function $r(v,t)$ equals to $\bar{r}$.
Therefore, differentiating both sides of equation (\ref{r}) w.r.t. $t$ we obtain
\begin{equation}
    \frac{d}{d t} r\left( v(\bar{r}, t),t \right)
  = \frac{\partial r}{\partial t}
       + \frac{\partial v}{\partial t} \frac{\partial r}{\partial v} = 0 \ .
\end{equation}
In the case of $\partial r / \partial v \ne 0$ we derive the time-evolution
equation for $v(\bar{r}, t)$ as
\begin{equation}
  \frac{\partial}{\partial t} v(\bar{r},t) =
 \frac{\partial}{\partial r}
    \Phi \left( \left(\frac{\partial v }{\partial r}\right)^{-1} \right) \ .
\end{equation}
In short, we compute a time-evolution of the pseudo-inverse function $v(\bar{r},t)$
for a given initial condition $\bar{r}$.
Since the computational domain is now restricted $r \in [0,1]$ we
can go around the above expanding support issue for any large time.\\
For the $\kappa$-diffusion equation, we set
\begin{equation}
 \Phi\big(p(v,t)\big) \equiv
p(v,t)\,\ku\big(p(v,t)\big) \ .
 \label{Fk}
\end{equation}
The top plot in figure \ref{typ-evo} shows a typical time-evolution of $p(v,t)$ with an initial state concentrated around $v=0$.
The bottom plot in the same figure \ref{typ-evo} shows the time-evolution 
for an initial state with double peaks. Note that the double peaks 
immediately disappear as time evolves and 
become a single peaked shape, which seems to be
asymptotically approaching to the $\kappa$-Gaussian.\\
We have run several simulations with different initial shapes $p(v,t=0)$
to confirm this asymptotic behavior.\\
A further indication on this asymptotic behavior can be obtained by studying the time evolution
of the function
\begin{equation}
  \eta(v,t)=\ln \left( \frac{p(v,t)}{p^{\rm tr}(v,t)} \right) \ ,
  \label{ratio}
\end{equation}
which is the logarithmic ratio between the numerical solution $p(v,t)$ of equation (\ref{kDE}) and the trial fitting function
\begin{equation}
  p^{\rm tr}(v,t)=\kexp\big(a(t)-b(t) v^2\big) \ .
  \label{k-Gau}
\end{equation}
This is a $\kappa$-Gaussian where $a(t)$ and $b(t)$ are the best fitted
parameters for the numerical solution at each time $t$.\\
In figure \ref{logratio}, we plotted the time evolution of $p(v,t)$ for
an initial $p(v,0)$ with a triangle shape. It is clear from this picture that $\eta(v,t)\to0$ as $t\to\infty$, i.e. the function (\ref{ratio}) gradually decreases to zero as time evolves, which gives a strong evidence that the numerical solution is asymptotically approaching to the $\kappa$-Gaussian function.

\subsection{Time dependent transformation}
In order to further confirm these asymptotic behaviors let us consider
a time-dependent transformation for the $\kappa$-generalized FPE \eqref{kFPE}, defined by
\begin{equation}
   w(v, t) = p\big( \xi(t), \tau(t) \big) \ ,
\label{t-transform}
\end{equation}
where
\begin{equation}
  \xi(t) = \frac{v}{a(t)} \ ,\quad \tau(t) = \ln a(t) \ ,
\end{equation}
and
\begin{equation}
a(t) = \left(1+ 2\,t \right)^{\frac{1}{2}} \ .
\end{equation}
Preliminarily, we observe that the transformation (\ref{t-transform}) maps 
a $\kappa$-Gaussian into a $\kappa$-Gaussian.\\
By transforming equation (\ref{kFPE}) according to equation (\ref{t-transform}) we obtain
\begin{equation}
  \frac{\partial }{\partial t} w(v, t) =
  D \frac{\partial^2}{\partial v^2}
   \left( \frac{ p(v,t)^{1+\kappa} + p(v,t)^{1-\kappa}}{2} \right)
  + \frac{w(v,t)}{2t + 1} \ ,\label{tr}
 \end{equation}
that is the $\kappa$-diffusive equation (\ref{kDE}) with the source term $w(v,t)/(2t+1)$.\\
In the large time limit, the source term becomes negligible and equation (\ref{tr}) can be well approximated by equation (\ref{kDE}). On the other hand, we recall that the solution of $\kappa$-generalized FPE \eqref{kFPE} approaches asymptotically to the stationary state which is
the $\kappa$-Gaussian \eqref{stationary}. Therefore, this solution is transformed, by means of equation (\ref{t-transform}), into another one that well approximates the solution of the $\kappa$-diffusive equation. In other words, this solution asymptotically approaches to a $\kappa$-Gaussian function.

%%%%%%%%%%%%%%%%%%%%%%%%%%%%%%%%%%%%%%%%%%%%%%%%%%%%%%%%%%%%%%%%%%%%%%%%%%%%%%%%%%%%%%%%

\section{Summary}
We have studied the asymptotic behaviors of the $\kappa$-diffusive equation,
which is naturally related with the $\kappa$-generalized statistical physics 
and the $\kappa$-FPE \eqref{kFPE}.
The analysis based on the standard Lie symmetry showed that the $\kappa$-Gaussian function is not a scale invariant solution
of the $\kappa$-diffusion equation. Notwithstanding, the $\kappa$-Gaussian 
plays a relevant r\^ole in the study of the time evolution of localized perturbations.
In fact, by performing several numerical simulations, with different
initial shapes, we have found strong evidence that these solutions, 
for large time, are well approximated by the $\kappa$-Gaussian. Arguments based on a time-dependent transformation performed on the $\kappa$-generalized FPE also support this result.\\
Further confirmations of this behavior could be obtained
by studying whether a $\kappa$-Gaussian is invariant under asymptotic
symmetries \cite{Gaeta} of the $\kappa$-diffusive equation.

\vspace{5mm}
\noindent{\bf Acknowledgment}\\
This research was partially supported by the Ministry of Education,
 Science, Sports and Culture (MEXT), Japan, Grant-in-Aid for
 Scientific Research (C), 19540391, 2008.


\begin{thebibliography}{99}
\bibitem{barotropic}
A barotropic fluid is one in which the pressure depends only on the density.
A common barotropic fluid used in astrophysics is a polytropic fluid, for which
the equation of state is written as $P = C \rho^{\gamma}$ with a constant $C$
and the polytropic index $\gamma$.
\end{thebibliography}
\end{document}